\documentclass[sigconf,natbib=true,anonymous=false]{acmart}

\usepackage{amssymb}
\usepackage{array}
\usepackage{url}
\usepackage{dblfloatfix}
\usepackage{booktabs}
\usepackage{amsmath}
\usepackage{tabularx}
\usepackage{colortbl}

\usepackage{graphics}
\usepackage{epsfig}
\usepackage{multirow}
\usepackage{flushend}

\usepackage{url}
\usepackage{tikz}
\usepackage{enumitem}
\usepackage{multicol}

\usepackage{verbatim}
\usepackage{float}
\usepackage{subcaption}

\usepackage{algorithm}
\usepackage{algpseudocode}

\usepackage{marginnote}
\usepackage{xspace}
\usepackage{balance}

\usepackage{todonotes}
\usepackage{xcolor}

\newcommand{\COMETS}{\texttt{CODEC}\xspace}

\newcommand{\squishlist}{
 \begin{list}{$\bullet$}
  { \setlength{\itemsep}{0pt}
     \setlength{\parsep}{1pt}
     \setlength{\topsep}{1pt}
     \setlength{\partopsep}{0pt}
     \setlength{\leftmargin}{1.5em}
     \setlength{\labelwidth}{1em}
     \setlength{\labelsep}{0.5em} } }

\newcommand{\squishend}{
  \end{list}  }



\copyrightyear{2022} 
\acmYear{2022} 
\setcopyright{acmlicensed}\acmConference[SIGIR '22]{Proceedings of the 45th International ACM SIGIR Conference on Research and Development in Information Retrieval}{July 11--15, 2022}{Madrid, Spain}
\acmBooktitle{Proceedings of the 45th International ACM SIGIR Conference on Research and Development in Information Retrieval (SIGIR '22), July 11--15, 2022, Madrid, Spain}
\acmPrice{15.00}
\acmDOI{10.1145/3477495.3531712}
\acmISBN{978-1-4503-8732-3/22/07}

\author{Iain Mackie}
\affiliation{
  \institution{University of Glasgow}
    \city{Glasgow, Scotland, UK}
}
\email{i.mackie.1@research.gla.ac.uk}

\author{Paul Owoicho}
\affiliation{
  \institution{University of Glasgow}
    \city{Glasgow, Scotland, UK}
}
\email{p.owoicho.1@research.gla.ac.uk}

\author{Carlos Gemmell}
\affiliation{
  \institution{University of Glasgow}
    \city{Glasgow, Scotland, UK}
}
\email{c.gemmell.1@research.gla.ac.uk}

\author{Sophie Fischer}
\affiliation{
  \institution{University of Glasgow}
    \city{Glasgow, Scotland, UK}
}
\email{sophie.fischer@glasgow.ac.uk}

\author{Sean MacAvaney}
\affiliation{
  \institution{University of Glasgow}
    \city{Glasgow, Scotland, UK}
}
\email{sean.macavaney@glasgow.ac.uk}

\author{Jeffrey Dalton}
\affiliation{
  \institution{University of Glasgow}
    \city{Glasgow, Scotland, UK}
}
\email{jeff.dalton@glasgow.ac.uk}

\renewcommand\footnotetextcopyrightpermission[1]{}

\begin{document}
\fancyhead{}

\title{CODEC: Complex Document and Entity Collection}

\begin{abstract}

% --- new abstract --- % 
\COMETS is a document and entity ranking benchmark that focuses on complex research topics.
We target essay-style information needs of social science researchers, i.e. `How has the UK's Open Banking Regulation benefited Challenger Banks?'.
\COMETS includes 42 topics developed by researchers and a new focused web corpus with semantic annotations including entity links.
This resource includes expert judgments on 17,509 documents and entities (416.9 per topic) from diverse automatic and interactive manual runs.
The manual runs include 387 query reformulations, providing data for query performance prediction and automatic rewriting evaluation.

\COMETS includes analysis of state-of-the-art systems, including dense retrieval and neural re-ranking. 
The results show the topics are challenging with headroom for document and entity ranking improvement. 
Query expansion with entity information shows significant gains in document ranking, demonstrating the resource's value for evaluating and improving entity-oriented search. 
We also show that the manual query reformulations significantly improve document ranking and entity ranking performance.
Overall, \COMETS provides challenging research topics to support the development and evaluation of entity-centric search methods. 

% --- new abstract --- % 

\end{abstract}

\maketitle
\section{Introduction}
\label{sec:intro}

Researchers spend considerable time exploring sources to understand key arguments, concepts and facts about a specific topic. 
For example, surveys show that many legal researchers, recruitment professionals, and healthcare researchers require high-recall Boolean or structured queries over domain-specific collections \cite{russell2018information}.
In contrast, \COMETS focuses on researchers within social sciences (History, Economics, and Politics) to develop complex topics that can be satisfied by web documents and entities within standard knowledge bases (i.e. Wikipedia).

\begin{figure}[h!]
    \centering
    \includegraphics[scale=0.25]{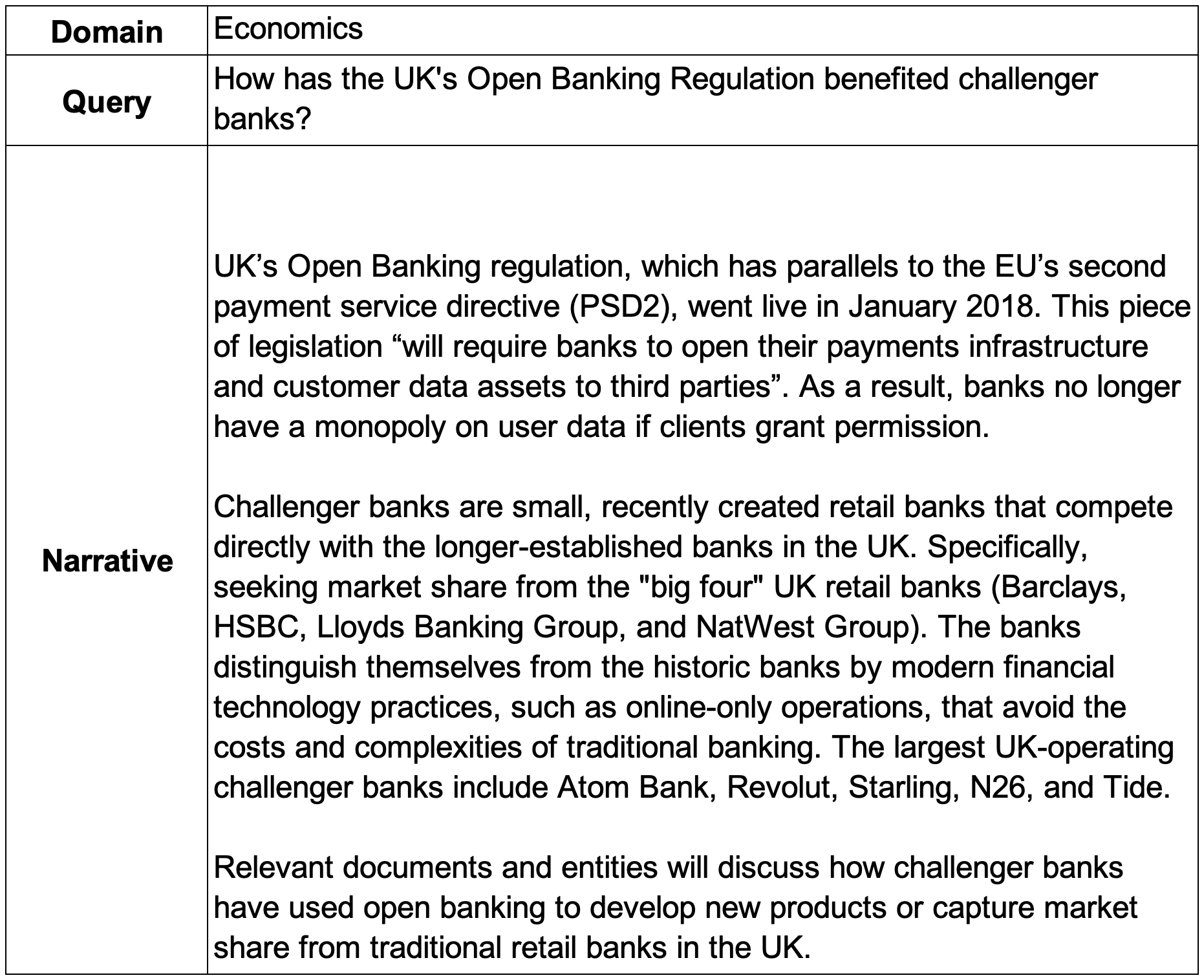}
    \caption{Example \COMETS topic: economics-1.}
    \label{fig:topic}
\end{figure}

Figure \ref{fig:topic} shows an example topic, where a financial researcher wants to understand \textit{How has the UK's Open Banking Regulation benefited Challenger Banks?}. The researcher would use a standard commercial search engine to identify valuable information, reformulating queries to investigate varying dimensions of the topic, e.g. `Open Banking products', `Challenger Banks fundraising', etc. 
Through this iterative process, the researcher is trying to identify useful sources (i.e. documents) and understand the critical concepts (i.e. entities). 
\COMETS \footnote{available at \url{https://github.com/grill-lab/CODEC} and \textit{ir-datasets} \cite{macavaney:sigir2021-irds}} is a dataset that seeks to benchmark document and entity ranking on complex long-form essay questions. 

\begin{figure*}[h!]
    \centering
    \includegraphics[scale=0.8]{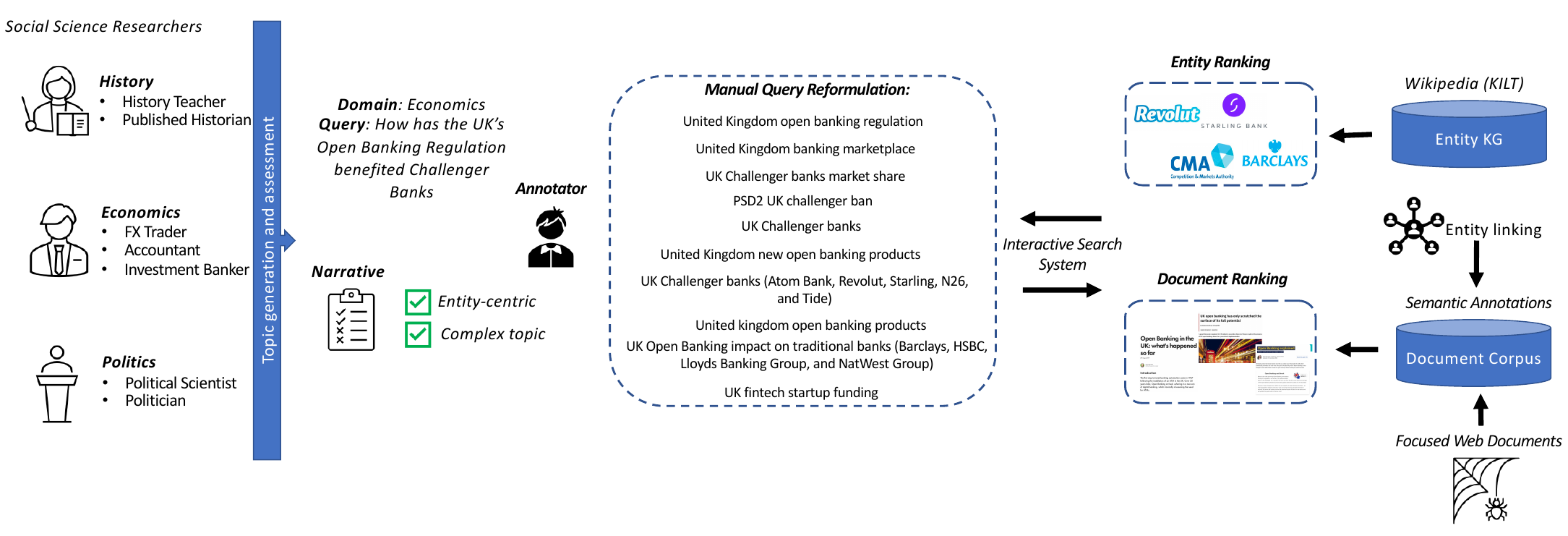}
    \caption{Overview of \COMETS. Social science researchers develop topics based on new complex topic criteria. Annotators assess initial pooled runs before using an interactive search system to issue manual queries and explore the topic. Documents are from \COMETS's focused web document collection and entities from KILT's \cite{petroni2021kilt} version of Wikipedia.} 
    \label{fig:overview}
\end{figure*}

Studies show that a large proportion of information needs are about entities or relate to entities \cite{kumar2010characterization, guo2009named}. 
This is particularly true for essay-style questions across social sciences, where the information need generally focuses on key events (e.g. \textit{How close did the world come to nuclear war during the Cuban Missile Crisis?}), people (e.g. \textit{How did Colin Kaepernick impact the political discourse about racism in the United States?}), or things (e.g. \textit{What technological challenges does Bitcoin face to becoming a widely used currency?}).
Previous work demonstrates that incorporating entity-based information improves ad-hoc retrieval, with the most notable improvements made on complex topics \cite{Dalton2014EntityQF}.

COmplex Document and Entity Collection (\COMETS) supports two distinct tasks: document ranking and entity ranking. 
Document ranking is the task, given an information need $Q$, to return a relevance-ranked list of documents $[D_1,...,D_N]$ from a document corpus $C_D$.
Entity ranking is the task, given an information need $Q$, to return a relevance-ranked list of entities $[E_1,..,E_N]$ from an entity knowledge base $KB_E$.
Entity links between the documents and entities provide structured connections between both tasks.
We also release the manual query reformulations from the researcher annotating the topic with mapped relevance judgments.
Figure \ref{fig:overview} shows the setup of these tasks.
Although tasks can be undertaken independently, \COMETS has aligned document and entity judgments to allow joint exploration of these tasks. 
This setup will allow researchers to leverage \COMETS to target two key limitations of current neural ranking models:  

\textbf{(1) Entity-centric representations:} Prior work has shown that ad-hoc retrieval \cite{Dalton2014EntityQF, xiong2017word} and entity ranking \cite{dietz2019ent} can be improved by leveraging entity information. 
Yet, current state-of-the-art methods rely solely on text representations and lack the understanding of entities and their relationships. 
For example, dense retrieval underperforms BM25 on even simple entity queries \cite{sciavolino2021simple}. 
\COMETS provides aligned document and entity judgments that are unified through entity-linking connections. 
For example, we demonstrate simple entity-based query expansion statistically improves MAP and Recall@1000 for document retrieval over strong initial retrieval systems. 
\COMETS enables entity-centric ranking models to be developed on complex topics.

\textbf{(2) Complex topics:}   
Compared to easier general factoid or short keyword queries \cite{huston2010evaluating}, \COMETS curates long natural-language queries where relevant topic information spans many entities and documents. 
Figure \ref{fig:topic} shows an example topic query and narrative.
\COMETS queries average 12.5 words in length, considerably longer than many datasets, i.e. TREC Deep Learning \cite{craswell2021overview} averages 5.8 words. 
We also deliberately do not release shorter keyword `title' queries, such as TREC CORE \cite{core2017trec} and TREC Robust \cite{Voorhees_TREC2004_robust}, to challenge end-to-end ranking on long, natural language queries.

We develop new complex topic criteria to produce topics that require deep knowledge and research to identify the key facts and arguments. 
Using topic narratives as a proxy for topic complexity, these contain on average 143.4 words and 23.7 explicitly mentioned entities. 
Annotators also required, on average, 9.2 manual query reformulations to reasonably understand these topics.

These query reformulations are released to allow the development of query expansion and query reformulation on complex topics. 
We show that query expansion using the query reformulations significantly improves MAP and Recall@1000 on document ranking and MAP, NDCG@10 and Recall@1000 on entity ranking. Additionally, we find the best manual reformulation performs better than the original query.

Figure \ref{fig:overview} shows the \COMETS dataset creation process that is designed for entity-centric ranking over complex topics. 
Social science experts across History (History teacher, History scholar), Economics (trader, accountant, investment banker) and Politics (political scientist, politician) generate 42 topics based on the developed criteria.

Experts also suggest 24 focused research websites for the corpus. 
Topics across History, Economics and Politics are selected because (1) this allows a  broad range of complex topics and (2) there is sufficient topic overlap to share a single web document corpus. 
After reviewing several standard document collections, none had enough focused content for these topics, with History topics having the poorest coverage. 
Thus, web content from Common Crawl is used with website-specific parsers to create a curated document corpus of around 700k heterogeneous web documents (blogs, news, interviews, etc.). 
Wikipedia via KILT \cite{petroni2021kilt} is the entity knowledge base (KB) for entity ranking.
Entity linking is run over the document corpus using REL \cite{vanHulst:2020:REL}, which is an effective end-to-end entity linker.

Experienced IR experts (the authors) produce 6,186 document (147.3 per topic) and 11,323 entity (269.6 per topic) relevance judgments. 
We use a two-stage assessment process: (1) assessing pooled runs of BM25 \cite{robertson1994some}, BM25 with RM3 expansion \cite{abdul2004umass}, ANCE \cite{xiong2020approximate}, MonoT5 \cite{nogueira2020document}, commercial search engine, and entity linkers \cite{Wu2020ZeroshotEL, decao2020autoregressive}. 
Then (2) allowing assessors to formulate a series of manual queries (average 9.2 queries per topic) to search and annotate different aspects of the topics.   
This assessment process is intended to simulate the research process of complex topics on document and entity ranking. 

We evaluate strong document and entity ranking systems using \COMETS, including sparse retrieval, dense retrieval, and neural Language Model (LM) re-ranking. 
The results show significant headroom for improvement within the current first-pass and re-ranking search systems on these complex topics. 
The best performing document ranking system has MAP under 0.35 and NDCG under 0.5, while the best-performing entity ranking system has MAP under 0.25 and NDCG under 0.45. 
Comparing these measures to comparable systems on TREC Deep Learning \cite{craswell2021overview}, where NDCG@10 is 0.7 and MAP is 0.55, highlights that \COMETS complex topics are challenging for current systems. 

Analysis shows a positive correlation between document relevance and the occurrence of the most relevant entities.
These findings support the complementary relationship between the document and entity ranking tasks and motivate the development of future models that leverage both \cite{Dalton2014EntityQF, dietz2019ent}.
We explore this directly by leveraging \COMETS entity judgments within document ranking via query expansion, which improves document ranking by a statistically significant amount. 

\COMETS is a valuable resource for IR researchers, supporting the development of new neural methods that leverage entity-centric representations. The key contributions of \COMETS are:

\begin{itemize}
    
    \item \textbf{Test collection:}
    We release a test collection to benchmark complex research topics on document and entity ranking.
    We produce new guidelines for complex topics and have social sciences experts generate 42 new topics. 
    We curate a new document corpus from 24 focused sources covering diverse social science domains across History, Economics, and Politics.
    We release 6,186 document and 11,323 entity judgments. 
    
    \item \textbf{Analysis of system performance:}
     We study the behaviour of strong systems (sparse retrieval, dense retrieval, and LM re-ranking) that highlights failures and provide headroom for improvement on complex topics.
     We highlight specific failures due to models lacking the ability to encode or utilize entities and their relationships. 

    \item \textbf{Aligned entity and document tasks:}
    We design the dataset to have aligned document and entity judgments to allow the development of new entity-centric ranking models. 
    We show that document relevance is positively correlated with the proportion of most relevant entities contained within the document.
    We show that we can improve document ranking using entity query expansion.
    
    \item \textbf{Query reformulation:}
    A two-stage assessment process allows the assessment of strong pooled runs, followed by a manual exploration of the topic using multiple live search systems. 
    These query reformulations (387 queries in total) are also released.
    We show that the best query reformulation outperforms the original query. 
    Additionally, query expansion that uses all query reformulations improves document and entity ranking. 

\end{itemize}

\section{Related Work}
\label{sec:related}

We provide an overview of the related literature across document ranking, entity ranking, and query reformulation. 

\subsection{Document ranking}

Document ranking is the task of retrieving a ranked list of relevant documents from a corpus given a specified information need.  
In recent years, pre-trained language models  \cite{Li_etal_2020_PARADE,  MacAvaney_etal_SIGIR2019, nogueira2020document} and dense retrieval systems \cite{khattab2020colbert, xiong2020approximate} have been shown to improve document ranking.
However, findings have shown failures of neural retrieval on even simple entity queries \cite{sciavolino2021simple}. 

Within the field of domain-specific research (i.e. legal, recruitment, and healthcare), high-recall Boolean or structured queries over domain-specific collections are standard \cite{russell2018information}.
\COMETS instead focuses on social sciences (History, Economics, and Politics) and open-ended essay-style topics that more closely align with web or newswire search.
For example, TREC Robust \cite{Voorhees_TREC2004_robust} and TREC CORE \cite{core2017trec}, provide a similar style of natural language queries. 
However, \COMETS topics are more current and provide extensive narratives, i.e. \COMETS narratives average 143.4 words versus TREC CORE's 44.0.

MS MARCO \cite{MS_MARCO_v1} is a family of passage and document test collections consisting of web queries, passages or documents, with sparse relevance judgments.
However, MS MARCO's annotation technique means that the queries tend to be artificially easy and exhibit undesirable qualities like the `maximum passage bias' \cite{Li_etal_2020_PARADE}.
TREC Deep Learning \cite{craswell2021overview} extends MS MARCO with dense judgments to provide a more useful benchmark, and DL-HARD \cite{mackie2021dlhard} develops a more challenging subset with annotations and metadata. 
\COMETS differs from these datasets in terms of length of queries, i.e. TREC Deep Learning averages 5.8 words compared to \COMETS's 12.5 words.
\COMETS provides a new focused document corpus that provides good coverage of complex social science topics.
Additionally, aligned entity and document judgments will allow researchers to explore the related task of entity ranking.

\subsection{Entity ranking}

Entity ranking is the task of retrieving a ranked list of relevant entities from an entity knowledge base given a specified information need. 
Past studies have shown that entity ranking improves by leveraging mentions in text passages to create a topic-specific text-entity graph \cite{dietz2019ent}.
Transformer-based embeddings have been shown to be a reasonable entity ranking baseline \cite{chatterjee2021entity}, with a strong performance on the related tasks such as entity linking \cite{wu2019zero}.
Entity ranking closely relates to entity aspect linking, where the task is to identify the fine-grained semantics of the entity that relates to a mention in a contextual passage \cite{nanni2018entity, ramsdell2020large}.
Incorporation of entity aspects has also been shown to improve entity ranking \cite{chatterjee2021entity}.

INEX 2009 XML Entity Ranking Track \cite{demartini2009overview} focuses on entity ranking and entity list completion from Wikipedia XML documents.
The queries are generally factoid in nature, i.e. `Italian Nobel prize winners' and `Formula 1 drivers that won the Monaco Grand Prix'.
In contrast, \COMETS asks entity ranking systems to rank important named entities or general concepts on complex topics, `What technological challenges does Bitcoin face to becoming a widely used currency?' Where relevant entities include [Cyberattack] and [Transaction Cost], as well as the explicitly mentioned [Bitcoin].    

DBpedia-Entity \cite{balog2013test} and DBpedia-Entity v2 \cite{hasibi2017dbpedia} are test collections for entity search over the structured DBpedia knowledge base.
These encompass four slightly different entity search tasks, i.e. named entity search, ad-hoc entity ranking, list completion, and natural language entity-based QA. 
The key difference is \COMETS uses a free-text based entity KG (Wikipedia), the topics are more open-ended, and there is an aligned document ranking task.

TREC CAR  \cite{dietz2017trec} is a passage and free-text entity ranking dataset built from Wikipedia and uses Wikipedia titles and headings as keyword queries. 
TREC CAR is the closest setup to \COMETS, and the sparse entity and document relevance judgements could provide a useful pre-training step. 
\COMETS differs based on the complex natural language queries, heterogeneous text corpus (instead of solely Wikipedia), and focus on document ranking. 

\subsection{Query Reformulation}

\citet{culpepper2021hard} highlight that users routinely reformulate queries to satisfy an information need, and show the high variance of retrieval performance across these query variants. 
In fact, \citet{culpepper2021hard} find that query formulations have a comparable effect to the actual topic complexity in terms of overall system performance.
\COMETS provides over 387 manual query reformulations to support research in this space, i.e. query performance prediction and automatic query reformulation.

\citet{liu2019comparative} show the benefits of human and automatic query reformulations on document ranking systems, with human reformulations being the most effective.
Analysis on \COMETS supports these findings that the best reformulations improve document and entity ranking.
ORCAS is a click dataset that aligns with TREC Deep Learning \cite{craswell2020orcas}, which is useful for identifying clusters of related queries or related documents. 
However, \COMETS provides manual query reformulation on a single information need, providing more fine-grained topic-specific query reformulations.

\section{CODEC}

\COMETS is a test collection that provides two tasks: document ranking and entity ranking. 
A complex topic is defined as an essay-style question where essential information can span across multiple entities and documents (see Section \ref{sec:topics} for more detail).

This dataset benchmarks a researcher who is attempting to find supporting entities and documents that will form the basis of a long-form essay discussing the topic from various perspectives. 
The researcher would explore the topic to (1) identify relevant sources and (2) understand key concepts. 

\subsection{Task Definition} 
\label{sec:task-def}

\COMETS supports both document ranking and entity ranking tasks. 
Document ranking systems have to return a relevance-ranked list of documents $[D_1,...,D_N]$, from a document corpus $C_D$, for a given natural language query $Q$. 
Entity ranking systems have to return a relevance-ranked list of entities $[E_1,...,E_N]$, from an entity knowledge base $KB_E$, for a given natural language query $Q$. 
Document ranking uses \COMETS's new document corpus and entity ranking uses KILT as the entity KB.
For the experimental setup, we provide four pre-defined `standard' folds for k-fold cross-validation to allow parameter tuning. 
Initial retrieval or re-ranking of provided baseline runs can both be evaluated using this test collection.  
\COMETS setup encourages exploration of the related entity and document ranking tasks; however, both tasks can also be undertaken in isolation.

\subsection{Topic Generation}
\label{sec:topics}

\COMETS provides complex topics that intend to benchmark the role of a researcher. 
Understanding these topics requires deep knowledge and investigation to identify the key documents and entities.

Social science experts from History (History teacher, published History scholar), Economics (FX trader, accountant, investment banker) and Politics (political scientists, politician) helped to generate interesting and factually-grounded topics. 
The authors develop the following criteria for complex topics: 

\begin{itemize}

    \item \textbf{Open-ended essay-style}: 
    Satisfying the information need of this topic comprehensively would require a long-form essay-style response. 
    Factoid questions or questions that only require a short answer are not suitable. 
    
    \item \textbf{Natural language question}: 
    The query should be long, natural language-based. 
    Keyword queries are not suitable.
    
    \item \textbf{Multiple points of view}: 
    It is preferable if the topic elicits debate and multiple points of view. 
    A good response would thus require an understanding of each of these dimensions.
    
    \item \textbf{Concern multiple key entities}: 
    It is preferable if the topic concerns multiple key entities (people, things, events, etc.). 
    
    \item \textbf{Complexity}: 
    It is preferable if the topic requires an educated adult to undertake significant research to understand it. 
    
    \item \textbf{Knowledge}: 
    It is preferable if the topic requires deep knowledge to understand satisfactorily. 
    
\end{itemize}

The domain experts write 42 topics with minimal post-processing from the authors to align styles or correct spelling or grammatical errors. 
There is an equal number of topics per target domain, i.e. 14 History topics, 14 Economics topics, and 14 Politics topics.
Each topic contains a query and narrative.
The query is the question the researcher seeks to understand by exploring documents and entities, i.e., the text input posed to the search system. 
The narratives provide an overview of the topic (key concepts, arguments, facts, etc.) and allow non-domain experts to understand the topic. 
Due to the complexity of these topics, the narratives are not completely comprehensive but provide a useful starting point for annotators. 
We also review pooled runs to assess whether topics are too easy (i.e. lots of highly ranked relevant documents) or do not align with the corpora (i.e. not enough relevant documents or entities to satisfy the information need).

Table \ref{tab:topic-stats} shows the average number of words and entities in topic queries and narratives.
Entity statistics are calculated by running GENRE \cite{decao2020autoregressive} over the queries and narratives.
An average of 12.5 words and 2.4 entities per query supports long natural language queries that include entities.
Narratives provide a good proxy for the complexity of the underlying information need, and 143.4 words and 23.7 entities support this complexity.

\begin{table}[h!]
\caption{Topic Statistics across 42 \COMETS topics.}
\label{tab:topic-stats}
\begin{tabular}{l|r|r|}
\cline{2-3}
                                                      & \multicolumn{1}{l|}{\textbf{Total}} & \multicolumn{1}{l|}{\textbf{Avg. Length}} \\ \hline
\multicolumn{1}{|l|}{\textbf{Query (Words)}}    & 524                                        & 12.5                                    \\ \hline
\multicolumn{1}{|l|}{\textbf{Query (Entities)}} & 102                                        & 2.4                                     \\ \hline
\multicolumn{1}{|l|}{\textbf{Narrative (Words)}}     & 6,021                                      & 143.4                                 \\ \hline
\multicolumn{1}{|l|}{\textbf{Narrative (Entities)}}  & 994                                        & 23.7                                  \\ \hline
\end{tabular}
\end{table}

\subsection{Document Corpus}
\label{sec:doc-corpus}

We want \COMETS document corpus to have enough high-quality coverage of current social science topics.

\subsubsection{Focused Content}

We perform initial exploration on standard document collections (MS MARCO, TREC Washington Post, etc.) with \COMETS topics but find critical coverage gaps within required research content.  
History topics have particularly low coverage and would require augmentation from historical authority sites.
This motivates building upon a subset of Common Crawl to create a new focused document corpus for the target domains.
Table \ref{tab:corpus_docs} shows the distribution of documents.

We leverage our domain experts, who recommend suitable seed websites or sections of websites.
The pool contains a mixture of clearly specialized websites (i.e. economicsdiscussion.net, history.com, brookings.edu) and several general newswire websites (bbc.co.uk, latimes.com, etc.).
Social science experts requested up-to-date newswire websites for contextualizing current economic and political topics. 
We also run the topics through a commercial search engine to ensure appropriate coverage and that each domain has enough representation.

\begin{table}[h!]
\caption{Distribution of Top 15 Websites in Document Corpus.}
\label{tab:corpus_docs}
\begin{tabular}{l|r|}
\cline{2-2}
                                                & \multicolumn{1}{l|}{\textbf{Count}} \\ \hline
\multicolumn{1}{|l|}{reuters.com}               & 172,127                                      \\ \hline
\multicolumn{1}{|l|}{forbes.com}                & 147,399                                      \\ \hline
\multicolumn{1}{|l|}{cnbc.com}              & 100,842                                      \\ \hline
\multicolumn{1}{|l|}{britannica.com}            & 93,484                                       \\ \hline
\multicolumn{1}{|l|}{latimes.com}               & 88,486                                       \\ \hline
\multicolumn{1}{|l|}{usatoday.com}              & 31,803                                       \\ \hline
\multicolumn{1}{|l|}{investopedia.com}          & 21,459                                       \\ \hline
\multicolumn{1}{|l|}{bbc.co.uk}                 & 21,414                                       \\ \hline
\multicolumn{1}{|l|}{history.state.gov}         & 9,187                                        \\ \hline
\multicolumn{1}{|l|}{brookings.edu}             & 9,058                                        \\ \hline
\multicolumn{1}{|l|}{ehistory.osu.edu}          & 8,805                                        \\ \hline
\multicolumn{1}{|l|}{history.com}               & 6,749                                        \\ \hline
\multicolumn{1}{|l|}{spartacus-educational.com} & 3,904                                        \\ \hline
\multicolumn{1}{|l|}{historynet.com}            & 3,811                                        \\ \hline
\multicolumn{1}{|l|}{historyhit.com}            & 3,173                                        \\ \hline
% \multicolumn{1}{|l|}{historynewsnetwork.org}    & 3,112                                        \\ \hline
% \multicolumn{1}{|l|}{smithsonianmag.com}        & 2,085                                        \\ \hline
% \multicolumn{1}{|l|}{factcheck.org}             & 1,084                                        \\ \hline
% \multicolumn{1}{|l|}{timemaps.com}              & 766                                          \\ \hline
% \multicolumn{1}{|l|}{history.hanover.edu}       & 333                                          \\ \hline
% \multicolumn{1}{|l|}{economicsdiscussion.net}   & 286                                          \\ \hline
% \multicolumn{1}{|l|}{military-history.org}      & 220                                          \\ \hline
% \multicolumn{1}{|l|}{digitalhistory.uh.edu}     & 162                                          \\ \hline
% \multicolumn{1}{|l|}{historyandpolicy.org}      & 75                                           \\ \hline
\hline
  \hline
\multicolumn{1}{|l|}{\textbf{TOTAL}}            & \textbf{729,824}                             \\ \hline
\end{tabular}
\end{table}

\subsubsection{Corpus Generation Pipeline}

The document corpus pipeline takes the focused seed websites and uses Common Crawl and URL pattern matching to extract ~300GB of HTML across recent crawls in 2021  (CC-MAIN-2021-[21,17,10,14]). 
We develop 24 custom BeautifulSoup HTML parsers to extract text and metadata while removing any advertising and formatting. 
This creates documents with fields:

\begin{itemize}
    \item \textbf{id}: Unique identifier is the MD5 hash of URL.
        
    \item \textbf{url}: Location of the webpage (URL).   
        
    \item \textbf{title}: Title of the webpage if available.
        
   \item \textbf{contents}: The text content of the webpage after removing any unnecessary advertising or formatting. 
   New lines provide some structure between the extracted sections of the webpage, while still easy for neural systems to process.   
\end{itemize}

We then run multiple filtering stages to ensure the documents are of suitable length and unique.
First, the extracted text has to contain at least 30 words, approximately a paragraph.
Second, we identify that several websites contain the same (or very similar) webpage hosted on different URLs. 
Thus, we run a de-duplication step by grouping webpages from the same website that (1) have the same \textit{title} and (2) cosine similarity between document tokens greater than 95\%.
We solely include the document with the longest \textit{contents} in the final corpus.
This removes 96,900 duplicates and results in a final corpus containing 729,824 documents.
The corpus is released in jsonlines format. 

\subsubsection{Entity Linking}

We run the REL \cite{vanHulst:2020:REL} entity linker over the entire 729,824 document corpus to provide structured connections between documents and entities.
REL is a lightweight neural entity linker that allows easy deployment and strong performance.
We use the suggested setup for mention detection, i.e. Flair \cite{akbik2018contextual} which is a Named Entity Recognition (NER) model based on contextualized word embeddings. 
We use REL's pre-trained model for candidate selection that uses a 2019-07 version of Wikipedia (i.e. closely aligns with the 2019/08/01 Wikipedia version for entity KB).   
For each document we provide a list of entity links containing fields:

\begin{itemize}
    \item \textbf{mention}: Text spans in document that is linked to entity.
    
    \item \textbf{prediction}: Top predicted entity link (Wikipedia title).

    \item \textbf{prediction\_kilt}: We map \textit{prediction} entity link to KILT id to align with entity KB and entity judgments.

    \item \textbf{candidates}: Top-k entity link candidates (Wikipedia title).

    \item \textbf{candidates\_kilt}: We map \textit{candidates} entity links to KILT ids to align with entity KB and entity judgments.

   \item \textbf{conf\_ed}: Score of Flair NER model.
   
   \item \textbf{score}: Scores of REL candidate selection model.

\end{itemize}

We release the full 18GB of entity links in jsonlines format.
This will allow researchers to use entity links within document and entity ranking easily. 
Table \ref{tab:corpus_els} shows breakdown of the 27.5m entity links (37.7/document) and 144.1m entity candidates (197.5/document).

\begin{table}[h!]
\caption{Entity Links on Document Corpus.}
\label{tab:corpus_els}
\begin{tabular}{l|r|r|}
\cline{2-3}
                                        & \multicolumn{1}{l|}{\textbf{Corpus Total}} & \multicolumn{1}{l|}{\textbf{Document Mean}} \\ \hline
\multicolumn{1}{|l|}{Entity Links}      & 27,482,650                                          & 37.7                                           \\ \hline
\multicolumn{1}{|l|}{Entity Candidates} & 144,127,482                                          & 197.5                                           \\ \hline
\end{tabular}
\end{table}

\subsection{Entity KB}
\label{sec:entity-kb}

\COMETS uses KILT's \cite{petroni2021kilt} Wikipedia KB for the entity ranking task, which is based on the 2019/08/01 Wikipedia snapshot.
KILT contains 5.9M preprocessed articles which are freely available to use.
The entity pages are primarily text-based with minimal structure to indicate headings or passages, i.e. very similar to Document Corpus.
KILT is selected for \COMETS's KB because it aligns with related knowledge-grounded tasks (fact-checking, open-domain QA, entity linking, etc.).
KILT also provides inter-entity entity links based on Wikipedia mentions, which could be helpful when identifying how related entities are to each other. 

\subsection{Relevance Criteria}
\label{sec:relevance}

We perform relevance assessment on a graded scale (between 0 and 3) using developed guidelines to ensure a consistent assessment process. 
Guidelines take inspiration from those of HC4 \cite{ecir2022hc4} and are adapted for our tasks (full guidelines online). 

\subsubsection{Document Criteria} 

The key question for document relevance is: \textit{How valuable is the most important information in this document?}

\begin{itemize}

    \item \textbf{Very Valuable (3):}  
    The most valuable information in the document would be found in the lead paragraph of a report written on the topic. 
    This includes central topic-specific arguments, evidence, or knowledge. 
    This does not include general definitions or background.  
  
    \item \textbf{Somewhat valuable (2):}  
    The most valuable information in the document would be found in the body of such a report. 
    This includes valuable topic-specific arguments, evidence, or knowledge.   
    
    \item \textbf{Not Valuable (1):} 
    Although on topic, the information contained in the document might only be included in a report footnote or omitted entirely. 
    This consists of definitions or background information.

    \item \textbf{Not Relevant (0):}  
    Not useful or on topic. 

\end{itemize}

\subsubsection{Entity Criteria}

The key question for entity relevance is: \textit{How valuable is understanding this entity to contextualize document knowledge?}

\begin{itemize}

     \item \textbf{Very Valuable (3):}  
    This entity would be found in the lead paragraph of a report written on the topic. 
    It is absolutely critical to understand what this entity is for understanding this topic. 
  
    \item \textbf{Somewhat valuable (2):}  
    The entity would be found in the body of such a report.
    It is important to understand what this entity is for understanding this topic.  
  
    \item \textbf{Not Valuable (1):} 
    Although on topic, this entity might only be included in a report footnote or omitted entirely. 
    It is useful to understand what this entity is for understanding this topic.  

    \item \textbf{Not Relevant (0):}  
    This entity is not useful or on topic. 

\end{itemize}

\subsection{Assessment Process}
\label{sec:assessment}

\COMETS uses a 2-stage assessment approach to balance adequate coverage of current systems while allowing annotators to explore topics using iterative search systems. 

\subsubsection{Initial Run Assessment}

We generate pools from runs using state-of-the-art sparse and dense retrieval methods. For document runs we use top-100 BM25 \cite{robertson1994some}, BM25 using RM3 expansion \cite{abdul2004umass}, ANCE \cite{xiong2020approximate}, BM25 re-ranked with MonoT5 \cite{nogueira2020document}, BM25 using RM3 expansion re-ranked with MonoT5, and ANCE re-ranked by MonoT5. 
We also use a commercial search engine where the top-100 search results are limited to the 24 corpus websites, and the URLs are mapped back to document ids.
Pyserini \cite{lin2021pyserini} is used for BM25 and BM25 with RM3 expansion with default parameters.
We use MS Marco fine-tuned versions of ANCE and MonoT5. 

For entity runs, we also use a pool of the top-100 results from BM25, BM25 using RM3 expansion, ANCE, BM25 re-ranked with MonoT5, BM25 with RM3 re-ranked with MonoT5, and ANCE re-ranked with MonoT5. 
We use ELQ \cite{Wu2020ZeroshotEL}, which is an end-to-end entity linking model for questions, to produce an entity run on the queries.
GENRE \cite{decao2020autoregressive}, sequence-to-sequence entity linking model, is used to produce an entity run using the narrative. 
We again use a commercial search engine where top-100 search results are limited to Wikipedia and URLs mapped back to document ids. 

We devise a weighting ratio for document and entity pooling based on an analysis of several topics across domains.  
This process takes (1) top-k for each initial system run, then (2) intersection across specified sub-groups, before (3) sampling until the required threshold is reached.
The pooling method provides an initial 60 documents and entities for annotators to assess, which provides a reasonable starting point for annotation before the topic exploration stage. 

Experienced IR annotators (the authors) judge the top 60 documents before doing the same for the top 60 entities.
Documents are deliberately judged before entities to provide the annotator with the necessary topic knowledge to assess entity relevance.  

\subsubsection{Topic Exploration}

After the initial runs are assessed, annotators are allotted between two and three hours to use live search systems to explore key dimensions of topics to find relevant documents or entities.
Annotators need to construct a minimum of 6 new manual query reformulations.
Figure \ref{fig:overview} shows the query reformulations for the economics-1 topic.
Annotators are encouraged to run these queries through a commercial search engine for spell checking and evaluate whether the results are on topic.  

The live search systems use a hybrid BM25, BM25 with RM3 expansion and ANCE for initial retrieval, with re-ranking from MonoT5. 
This system returns the top 50 documents and top 50 entities to the assessor.
Similar to how a researcher would use commercial search systems to explore a topic iteratively, annotators do not need to assess all returned documents and entities. 
Annotators are encouraged to scan returned result lists using the title and keyword highlighting to decide whether the document or entity is worth considering before annotating. 
This process is designed to identify the highly-relevant documents and entities not currently returned by baseline systems.
Annotators are encouraged to keep searching until they cannot find new relevant documents or entities.

\subsubsection{Judgments}

Table \ref{tab:judgments} shows the distribution of judgments across the 42 judged topics, which includes 6,186 document judgments (147.3 per topic) and 11,323 entity judgments (269.6 per topic).
\textit{Highly Valuable (3)} only makes up 7\% of document judgments and 7\% of entity judgments. 
\COMETS also releases the manual query reformulations, with the topic exploration phase providing around 74\% of overall judgements.
There are 387 additionally issued queries overall (9.2 per topic), which can be used to explore query performance prediction or system improvement via query reformulations.

\begin{table}[h!]
\caption{Judgment distribution across 42 topics.}
\label{tab:judgments}
\begin{tabular}{|r|r|r|}
\hline
\multicolumn{1}{|l|}{\textbf{Judgment}} & \multicolumn{1}{l|}{\textbf{Document Ranking}} & \multicolumn{1}{l|}{\textbf{Entity Ranking}} \\ \hline
\textbf{0}                              & 2,353                                          & 7,053                                        \\ \hline
\textbf{1}                              & 2,210                                          & 2,241                                        \\ \hline
\textbf{2}                              & 1,207                                            & 1,252                                        \\ \hline
\textbf{3}                              & 416                                            & 777                                          \\ \hline
\multicolumn{1}{|l|}{\textbf{TOTAL}}    & \textbf{6,186}                                 & \textbf{11,323}                              \\ \hline
\end{tabular}
\end{table}

\subsubsection{Evaluation} 
\label{sec:evaluation}

We provide TREC-style query-relevance files with graded relevance judgments (0-3) for entity and document evaluation.
The official measures for both tasks include MAP and Recall@1000 with binary relevance above 1 (i.e. relevance mappings: \textit{0=0.0, 1=0.0, 2=1.0, 3=1.0}), and NDCG@10 with custom weighted relevance judgments (i.e. relevance mappings: \textit{0=0.0, 1=0.0, 2=1.0, 3=2.0}). 
We deliberately gear measures toward the most key documents and entities (i.e. relevance scores of 2 or 3) to prioritise systems ranking these higher vs more tangential but on-topic information (i.e. relevance score of 1).

MAP assumes the user wants to find many relevant documents or entities, exposing ranking order throughout the run. 
On the other hand, NDCG10 with custom scaling to overweight critical information aim to provide a clear signal of whether systems highly rank the essential documents and entities.  
Due to recall being important for research-based tasks, Recall@1000 shows missed information.

\section{Experimental Results}
\label{sec:exp}

We conduct an in-depth analysis of sparse, dense and neural re-ranking systems on \COMETS across document and entity tasks.
Document ranking shows a neural re-ranker with query expansion is the best performing system, and entity ranking is particularly challenging for neural systems in a zero-shot setting.
We highlight critical system failures, including models lacking (1) the ability to filter based on entities and relationships and (2) identify latent dimensions of the topic.
Using \COMETS's aligned document and entity judgments, we show that an entity-based query expansion technique significantly outperforms other systems.
We also demonstrate how manual query reformulations can improve system performance.
Firstly, showing the best query reformulation significantly outperforms the original query.
Secondly, we demonstrate that a reformulation-based query expansion technique significantly outperforms other systems.

\subsection{Systems}
\label{sec:baselines}

For sparse retrieval methods, the full text of entities and documents are both indexed using Pyserini \cite{lin2021pyserini}, with Porter stemming, and stopwords removed.
We use the released `standard' four folds for cross-validation on sparse baselines and release the tuned parameters for each fold. 
We optimise \textbf{BM25} \cite{robertson1994some} for MAP via parameter grid search of $k1$ (between 0.1 and 5.0 with step of 0.2) and $b$ (between 0.1 and 1.0 with step of 0.1).

For \textbf{BM25+RM3} \cite{abdul2004umass}, we use tuned $k1$ and $b$ fold parameters for BM25, and optimise RM3 for MAP via parameter grid search of $fb\_terms$ (between 5 and 95 with step of 5), $fb\_docs$ (between 5 and 20 with step of 5), and $original\_query\_weight$ (between 0.2 and 0.8 with step of 0.1).
    
ANCE \cite{xiong2020approximate} is a dense retrieval model that constructs harder negative samples using the Approximate Nearest Neighbor (ANN) index.
We use an MS Marco fined-tune ANCE model and Pyserini's wrapper for easy indexing.
Following the methodology in the ANCE paper, \textbf{ANCE+FirstP} takes the first 512 BERT tokens of each document to represent that document.
While \textbf{ANCE+MaxP} shards the document into a maximum of four 512-token shards with no overlap, and the highest-scoring shard represents the document.   
For entity ranking, we solely used ANCE+FirstP due to computational overhead.
Using the first paragraph of Wikipedia to represent an entity is common practice in entity linking \cite{Wu2020ZeroshotEL}.

\textbf{T5} \cite{nogueira2020document} is state-of-the-art LM re-ranker that casts text re-ranking into a sequence-to-sequence setting and has shown impressive results.
We use Pygaggle's \cite{lin2021pyserini} MonoT5 model, which is fine-tuned using MS Marco.
The model is not fine-tuned specifically on \COMETS and is used in a transfer-learning setup because of the size and scope of the current benchmark.
For document and entity ranking, we employ a max-passage approach similar to \citet{nogueira2020document} to re-rank initial retrieval runs (BM25, BM25+RM3, ANCE-FirstP, ANCE-MaxP).
The document is sharded in 512 tokens shards with a 256 overlapping token window (maximum 12 shards per document), and the highest scored shard is taken to represent the document.

Significance testing is conducted using a paired-t-test approach at a 5\% threshold, which is common within the IR community \cite{smucker2007comparison}.

\subsection{Analysis of Current Systems}
\label{sec:analysis-current}

The official evaluation measures are calculated on runs to a depth of 1,000 documents and entities (full result tables in Github repository).
\COMETS performs all evaluation using the \textit{ir\_measures} package \cite{macavaney:ecir2022-irm} and provides commands to make evaluation straightforward.

\begin{table}[h!]
\caption{Document ranking performance. $Bold$ indicates best system and $(^{\triangle})$ indicates 5\% paired-t-test significance against BM25.}
\label{tab:document-baselines}
\begin{tabular}{l|r|r|r|}
\cline{2-4}
                                              & \multicolumn{1}{l|}{\textbf{MAP}} & \multicolumn{1}{l|}{\textbf{NDCG@10}} & \multicolumn{1}{l|}{\textbf{Recall@1000}} \\ \hline
\multicolumn{1}{|l|}{\textbf{BM25}}           & 0.213                             & 0.322                                 & 0.762                                     \\ \hline
\multicolumn{1}{|l|}{\textbf{BM25+RM3}}       & $0.233^{\triangle}$               & 0.327                                 & $\textbf{0.800}^{\triangle}$               \\ \hline
\multicolumn{1}{|l|}{\textbf{ANCE-MaxP}}      & 0.186                             & 0.363                                 & 0.689                                     \\ \hline
\multicolumn{1}{|l|}{\textbf{BM25+T5}}        &  $0.340^{\triangle}$              & $0.468^{\triangle}$                   & 0.762                                     \\ \hline
\multicolumn{1}{|l|}{\textbf{BM25+RM3+T5}}    & $\textbf{0.346}^{\triangle}$      & $0.472^{\triangle}$          & $\textbf{0.800}^{\triangle}$             \\ \hline
\multicolumn{1}{|l|}{\textbf{ANCE-MaxP+T5}}   & $0.316^{\triangle}$               & $\textbf{0.481}^{\triangle}$                   & 0.689                                     \\ \hline
\end{tabular}
\end{table}

Table \ref{tab:document-baselines} shows system performance for document ranking. 
Based on Recall@1000 of 0.800 the best performing method is BM25+RM3, outperforming BM25 and dense retrieval from ANCE-MaxP.
BM25+RM3 adds pseudo-relevant terms to the query based on a first-pass retrieval run; 80-95 terms are optimal.
For example, RM3 improves Recall@1000 on economics-18 topic, \textit{Was the crash that followed the dot-com bubble an overreaction considering the ultimate success of the internet?} by 32\% by adding or increasing the weight of terms such as [Amazon], [Pets.com], and [crash].
Many of these terms are entities, which supports research based on entity expansion. 

\subsubsection{Document Ranking}

An example of a hard topic for initial retrieval is economics-12, \textit{What are the common problems or criticisms aimed at public sector enterprises?}, with Recall@1000 of under 0.55 for all systems.
This topic requires a lot of latent knowledge, and analysis of relevant documents shows they contain minimal keyword overlap with the query.
This is supported by the annotator having to enter sixteen wide-ranging queries reformulations to find relevant documents and entities.

T5-MaxP improves all initial retrieval runs, with BM25+RM3+T5 having the highest MAP (0.346) and ANCE-MaxP+T5 having the highest NDCG@10 (0.481). 
However, an overall MAP of under 0.35 and NDCG@10 under 0.5 leave sufficient headroom for new document ranking systems to improve on complex queries. 
For comparison, TREC Deep Learning similar systems have an approximate MAP of 0.55 and NDCG@10 of 0.70.

For example, a hard topic of document re-ranking across all systems is politics-22, \textit{What was the role of technology in the Arab Spring?}. 
BM25+RM3+T5 has a Recall@1000 of 0.8 but an NDCG@10 of only 0.20. 
By analysing the top-ranked baseline runs, it is clear that models cannot filter documents based on `technology' (concept) used during the `Arab Spring' (event). 
For example, several top-ranked documents discuss Arab startups, science in Islamic World, or Blockchain in Politics.

\subsubsection{Entity Ranking}

\begin{table}[h!]
\caption{Entity ranking performance. $Bold$ indicates best system and $(^{\triangle})$ indicates 5\% paired-t-test significance against BM25.}
\label{tab:entity-baselines}
\begin{tabular}{l|r|r|r|}
\cline{2-4}
                                              & \multicolumn{1}{l|}{\textbf{MAP}} & \multicolumn{1}{l|}{\textbf{NDCG@10}} & \multicolumn{1}{l|}{\textbf{Recall@1000}} \\ \hline
\multicolumn{1}{|l|}{\textbf{BM25}}           & 0.181                             & 0.397                                 & 0.615                                     \\ \hline
\multicolumn{1}{|l|}{\textbf{BM25+RM3}}       & $\textbf{0.209}^{\triangle}$      & \textbf{0.412}                        & $\textbf{0.685}^{\triangle}$              \\ \hline
\multicolumn{1}{|l|}{\textbf{ANCE-FirstP}}    & 0.076                             & 0.269                                 & 0.340                                     \\ \hline
\multicolumn{1}{|l|}{\textbf{BM25+T5}}        & 0.172                             & 0.361                                 & 0.615                                     \\ \hline
\multicolumn{1}{|l|}{\textbf{BM25+RM3+T5}}    & 0.179                             & 0.362                                 & $\textbf{0.685}^{\triangle}$              \\ \hline
\multicolumn{1}{|l|}{\textbf{ANCE-FirstP+T5}} & 0.136                             & 0.407          & 0.340                                     \\ \hline
\end{tabular}
\end{table}

Table \ref{tab:entity-baselines} shows the system performance on the entity ranking task. 
Performance for entity ranking is lower when compared to document ranking, emphasising that entity ranking is a challenging task within the \COMETS setup.
The best system is BM25+RM3 with Recall@1000 of 0.685, which has a statistically significant difference compared with BM25. 
ANCE-FirstP's Recall@1000 of 0.340 is significantly worse than other initial retrieval methods.  
This could be partially driven by ANCE only using the first passage (and not the whole Wikipedia page) or entity ranking being different from MS Marco document ranking fine-tuning.

\begin{table}[h!]
\caption{Entity ranking performance. $Bold$ indicates best system and $(^{\triangle})$ indicates 5\% paired-t-test significance against BM25.}
\label{tab:entity-baselines}
\begin{tabular}{l|r|r|r|}
\cline{2-4}
                                              & \multicolumn{1}{l|}{\textbf{MAP}} & \multicolumn{1}{l|}{\textbf{NDCG@10}} & \multicolumn{1}{l|}{\textbf{Recall@1000}} \\ \hline
\multicolumn{1}{|l|}{\textbf{BM25}}           & 0.181                             & 0.397                                 & 0.615                                     \\ \hline
\multicolumn{1}{|l|}{\textbf{BM25+RM3}}       & $\textbf{0.209}^{\triangle}$      & \textbf{0.412}                        & $\textbf{0.685}^{\triangle}$              \\ \hline
\multicolumn{1}{|l|}{\textbf{ANCE-FirstP}}    & 0.076                             & 0.269                                 & 0.340                                     \\ \hline
\multicolumn{1}{|l|}{\textbf{BM25+T5}}        & 0.172                             & 0.361                                 & 0.615                                     \\ \hline
\multicolumn{1}{|l|}{\textbf{BM25+RM3+T5}}    & 0.179                             & 0.362                                 & $\textbf{0.685}^{\triangle}$              \\ \hline
\multicolumn{1}{|l|}{\textbf{ANCE-FirstP+T5}} & 0.136                             & 0.407          & 0.340                                     \\ \hline
\end{tabular}
\end{table}

T5 re-ranking zero-shot does not improve well-tuned sparse retrieval systems but improves ANCE-FirstP initial retrieval run.
The best end-to-end retrieval system is BM25 with RM3 expansion, with a MAP of 0.209 and an NDCG@10 of 0.412.

Analysis of system failures shows key concepts proved particularly hard to retrieve. 
For example, in topic economics-2, \textit{What technological challenges does Bitcoin face to becoming a widely used currency?}, all retrieval systems return anticipated named entities, i.e. [Bitcoin], [Blockchain], and [Satoshi Nakamoto]. 
However, systems miss the key concepts that are needed to truly understand this information need, i.e. [Transaction time], [Transaction cost], [Quantum technology], [Carbon footprint], and [Cyberattack]. 
Looking at the judgment mappings, these missed entities come from manual queries issued by the researcher looking at these specific dimensions.
This motivates improved representations of entities that incorporate entity language models based upon document mentions that are more query-specific.

Topics with core entities explicitly named in the query have better entity ranking performance. 
For example, history-17, \textit{How significant was Smallpox in the Spanish defeat of the Aztecs?}, all systems placed [History of smallpox in Mexico], [Fall of Tenochtitlan], and [Spanish conquest of the Aztec Empire] in top ranks.

\subsection{Entities in Document Ranking}
\label{sec:entity-link-exp}

\COMETS allows researchers to explore the role of entities in document ranking using the provided document judgment, entity judgments, and entity links that connect documents and entities. 

\begin{figure}[h!]
    \centering
    \includegraphics[scale=0.37]{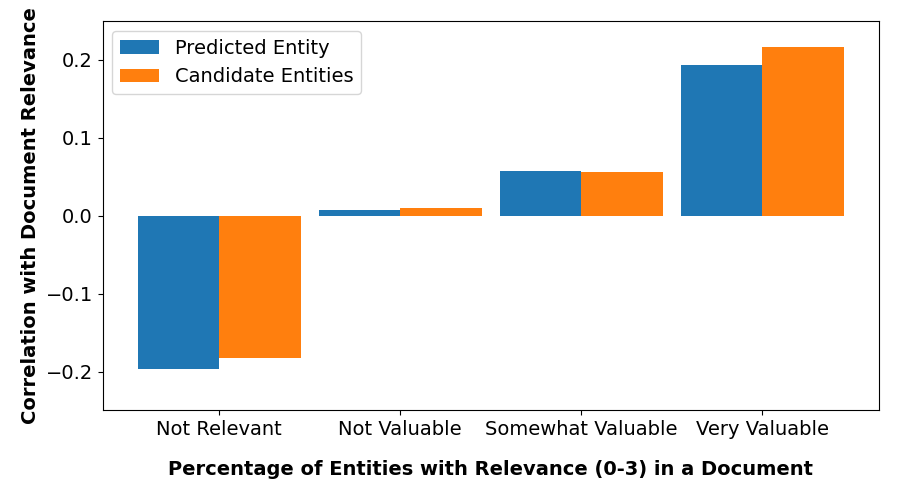}
    \caption{Correlation of document and entity relevance.}
    \label{fig:entity-to-doc-rank}
\end{figure}

To understand the relationship between document and entity relevance, we take the 6,186 document judgments and map the relevance of entities mentioned in each document using the entity links and entity judgments.
We assume entities without judgments are \textit{Not Relevant}.
We analyse both top-1 predicted entity and top-k candidate entities for document ranking.
Figure \ref{fig:entity-to-doc-rank} shows the Pearson Correlation Coefficient between document relevance and the percentage of entities in the document grouped by relevance, i.e. \textit{Not Relevant}, \textit{Not Valuable}, \textit{Somewhat Valuable},\textit{Very Valuable}. 
Both predicted entity (+0.19) and candidate entities (+0.22) support that documents with higher proportions of \textit{Very Valuable} entities are positively correlated with document relevance.

We develop an entity query feedback method to build on these findings.
Prior work has shown enriching queries with entity-based information improves ad hoc ranking performance \cite{Dalton2014EntityQF}.

\textbf{Entity-QE} is an oracle entity feedback method that enriches the query with names of  relevant entities taken from entity judgments.
We use Pyserini BM25 for initial retrieval, removing stopwords, using Porter stemming, and using \COMETS tuned BM25 fold parameters.
With \COMETS standard four-folds, we cross-validate the (1) weighting of original query terms, (2) weighting of \textit{Very Valuable} entity terms, and (3) weighting of \textit{Somewhat Valuable} terms. Across the four folds, Entity-QE term weighting is: (1) original queries average 9.2 terms with 80\% weighting, with \textit{Very Valuable} entities adding 42.6 terms on average with 16\% weighting, and \textit{Somewhat Valuable} entities adding 77.8 terms on average with 4\% weighting.

\begin{table}[h!]
\caption{Entity-QE Document ranking. $Bold$ indicates best system and $(^{\triangle})$ indicates 5\% paired-t-test significance against BM25+RM3+T5.}
\label{tab:entity-qe}
\begin{tabular}{l|r|r|r|}
\cline{2-4}
                                            & \multicolumn{1}{l|}{\textbf{MAP}} & \multicolumn{1}{l|}{\textbf{NDCG@10}} & \multicolumn{1}{l|}{\textbf{Recall@1000}} \\ \hline
\multicolumn{1}{|l|}{\textbf{BM25+RM3}}     & 0.223                             & 0.327                                 & 0.800                                      \\ \hline
\multicolumn{1}{|l|}{\textbf{Entity-QE}}    & 0.287                             & 0.405                                 & $\textbf{0.857}^{\triangle}$              \\ \hline
\multicolumn{1}{|l|}{\textbf{BM25+RM3+T5}}  & 0.346                             & 0.472                                 & 0.800                                      \\ \hline
\multicolumn{1}{|l|}{\textbf{Entity-QE+T5}} & $\textbf{0.356}^{\triangle}$                    & \textbf{0.476}                        & $\textbf{0.857}^{\triangle}$              \\ \hline
\end{tabular}
\end{table}

Table \ref{tab:entity-qe} shows Entity-QE improves Recall@1000 to 0.857, which is statistically significant when compared to the best initial retrieval systems BM25+RM3. 
\textbf{Entity-QE+T5} uses T5 as a re-ranker with the same setup as used in baseline runs and improves NDCG@10 to 0.476 and MAP to 0.356, a statistically significant improvement.
Overall, these findings support that entity-centric ranking methods benefit complex topics.
\COMETS having aligned document and entity judgments will enable new classes of neural ranking models to be developed and evaluated.

\subsection{Query Reformulation}

In this section, we study the utility of the manually reformulated queries used by the experts. 
We show that the best manual reformulation outperforms the original query on document and entity ranking. 
We also develop a query expansion method that uses all query reformulations that improve over the strong baselines.   

\subsubsection{Best Reformulation vs Original Query}

\begin{figure}[h!]
    \centering
    \includegraphics[scale=0.35]{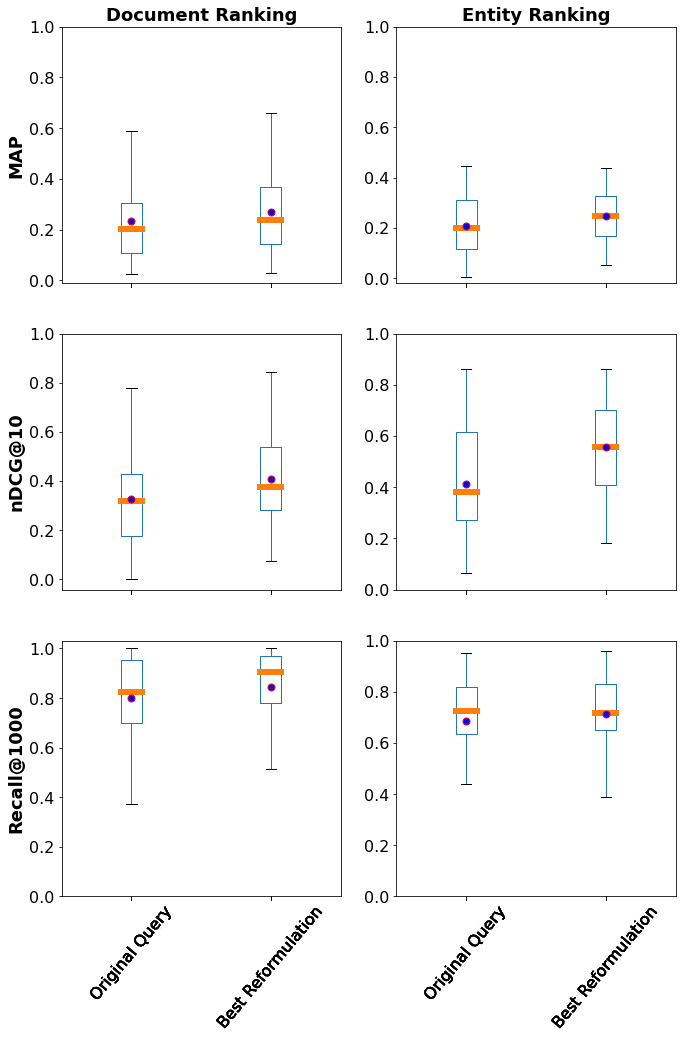}
    \caption{Boxplot BM25+RM3 topic performance of (1) original query, and (2) best manual query reformulation. Blue dot indicates means and orange line median across topics.}
    \label{fig:query-reform}
\end{figure}

We use a tuned BM25 and RM3 expansion model to analyse the performance of query reformulations against the original query, as this is a strong system across document and entity ranking. 
Figure \ref{fig:query-reform} shows the distribution of the best query reformulation against the original query across document and entity ranking for MAP, NDCG@10, and Recall@1000.

The best query reformulation improves Recall@1000 of document ranking to 0.845, a statistically significant difference. 
As is depicted in the boxplot, the best reformulation leads to almost 75\% of topics having Recall@1000 over 0.80. 
However, the best query reformulation has a smaller relative improvement on Recall@1000 for entity ranking (0.712) and is not statistically significant.
This suggests that several query reformulations are required for a robust initial entity ranking (as shown in Section \ref{sec:reform-eq}).

Analysing history-6 topic, \textit{What were the lasting social changes brought about by the Black Death?}, the original query performs poorly with a Recall@1000 of 0.428 on document ranking.
However, seven of thirteen query reformulations have Recall@1000 between 0.810 and 0.905, i.e \textit{The Black Death (Bubonic Plague) and end of Feudalism}, \textit{The Black Death (Bubonic Plague) and the Renaissance}, and \textit{Bubonic Plague / Black Death and Roman Catholic Church}.
The researcher is iterating on entity names and synonym expansion to identify missing documents and entities.

The best reformulation significantly improves document and entity ranking compared to the original query on MAP and NDCG@10 measures.
Document ranking MAP improves from 0.233 to 0.270, and NDCG@10 from 0.327 to 0.407. 
NDCG@10 saw the largest relative improvement, with around 75\% topics having an NDCG@10 over 0.3 (i.e. proportionally fewer failing topics than the original query).   
Similarly, the best query reformulation significantly improves entity ranking, with MAP improving from 0.209 to 0.248 and NDCG@10 from 0.412 to 0.557.

Entity ranking NDCG@10 saw a 35\% improvement due to the best query reformulation, the largest relative improvement of any measure across either task. 
Analysing the runs, this was driven by query reformulations accessing specific clusters of highly relevant entities within the top ranks. 
For example, the original query for topic history-15, \textit{Why did Winston Churchill lose the 1945 General Election after winning World War II?}, had an NDCG@10 of 0.323. 
The best query reformulation, \textit{Appeasement and Great Depression cost Conservatives in 1945 General Election}, improves NDCG@10 to 0.609. 
The improvement of top-ranked entities is due to the introduction of key events (i.e. [Appeasement] and [Great Depression]) and entities (i.e. [Conservatives]) being part of the query reformulation.

\subsubsection{Reformulation Query Expansion}
\label{sec:reform-eq}

We develop a query feedback method \textbf{Reform-QE}, which uses both the original and query reformulation terms.
We use BM25 in a similar setup to Entity-QE, cross-validating the weighting of the original query terms against the weighting of the aggregate query reformulation terms.
The original queries average 9.2 terms, and the aggregate query reformulation averages 42.8 terms.
For document ranking, the original queries average 66.7\% weighting and aggregate query reformulation averages 33.3\% weighting across the folds.
For entity ranking, the original queries average 60\% weighting and aggregate query reformulation averages 40\% weighting across the folds.
\textbf{Reform-QE+T5} uses T5 as a re-ranker with the same setup as used in baseline runs. 

Table \ref{tab:reform-qe-doc} depicts document ranking results for the query feedback methods that use the query reformulations.
Reform-QE significantly improves over the best initial retrieval system, BM25+RM3+T5, achieving Recall@1000 of 0.864.
Reform-QE+T5 also has the highest NDCG@10 and a statistically significant improvement in MAP when compared with BM25+RM3+T5.

\begin{table}[h]
\caption{Reform-QE Document ranking. $Bold$ indicates best system and $(^{\triangle})$ indicates 5\% paired-t-test significance against BM25+RM3+T5.}
\label{tab:reform-qe-doc}
\begin{tabular}{l|r|r|r|}
\cline{2-4}
                                            & \multicolumn{1}{l|}{\textbf{MAP}} & \multicolumn{1}{l|}{\textbf{NDCG@10}} & \multicolumn{1}{l|}{\textbf{Recall@1000}} \\ \hline
\multicolumn{1}{|l|}{\textbf{BM25+RM3}}     & 0.223                             & 0.327                                 & 0.800                                      \\ \hline
\multicolumn{1}{|l|}{\textbf{Reform-QE}}    & 0.275                             & 0.384                                 & $\textbf{0.864}^{\triangle}$              \\ \hline
\multicolumn{1}{|l|}{\textbf{BM25+RM3+T5}}  & 0.346                             & 0.472                                 & 0.800                                      \\ \hline
\multicolumn{1}{|l|}{\textbf{Reform-QE+T5}} & $\textbf{0.357}^{\triangle}$                    & \textbf{0.474}                        & $\textbf{0.864}^{\triangle}$              \\ \hline
\end{tabular}
\end{table}

Table \ref{tab:reform-qe-ent} shows entity ranking results of the query feedback methods that use the query reformulations. 
Reform-QE significantly outperforms the best entity system, BM25+RM3, across MAP, NDCG@10, and Recall@1000.
There is a larger relative improvement when using query reformulations compared to document ranking, highlighting how several queries are needed to expose the full range of relevant entities.

\begin{table}[h]
\caption{Reform-QE Entity ranking. $Bold$ indicates best system and $(^{\triangle})$ indicates 5\% paired-t-test significance against BM25+RM3.}
\label{tab:reform-qe-ent}
\begin{tabular}{l|r|r|r|}
\cline{2-4}
                                            & \multicolumn{1}{l|}{\textbf{MAP}} & \multicolumn{1}{l|}{\textbf{NDCG@10}} & \multicolumn{1}{l|}{\textbf{Recall@1000}} \\ \hline
\multicolumn{1}{|l|}{\textbf{BM25+RM3}}     & 0.209                             & 0.412                                 & 0.685                                     \\ \hline
\multicolumn{1}{|l|}{\textbf{Reform-QE}}    & $\textbf{0.253}^{\triangle}$      & $\textbf{0.525}^{\triangle}$          & $\textbf{0.738}^{\triangle}$              \\ \hline
\end{tabular}
\end{table}

Overall, query reformulations offer systems a chance to explore complex topics and access information about key dimensions of the topic not explicitly expressed in the query.
\COMETS query reformulations allow research into query reformulation or query performance prediction on complex topics. 

\section{Conclusion}
\label{sec:con}

We introduce \COMETS, a document and entity ranking resource that focuses on complex research topics.
Social science researchers produce 42 topics spanning History, Economics, and Politics.
To support open research, we create a new semantically annotated focused collection derived from subsets of the Common Crawl. 
\COMETS is grounded to the KILT's Wikipedia knowledge base for entity linking and retrieval.
We provide 17,509 document and entity judgments (416.9 per topic) by assessing the pooled initial runs and manual exploration of the topics using interactive search systems, adding 387 manual query reformulations (9.2 per topic).

\COMETS system analysis demonstrates topics are challenging for state-of-the-art traditional models and neural rankers.
Failures demonstrate encoding entities and relationships is challenging for both document and entity ranking. 
Specifically, queries with large amounts of latent knowledge, where new expansion techniques are a promising research direction.

We find that document relevance is positively correlated with the occurrence of relevant entities.
We leverage this relationship with an entity query expansion method that outperforms strong baseline systems on document ranking.
We also demonstrate that query reformulation can play an important role in accessing latent dimensions within complex topics.
Both individual query reformulations and aggregated reformulations improve document and entity ranking.  
Overall, this resource represents an important step toward developing and evaluating entity-centric search models on complex topics.

% \newpage

\section{Future Work}
\label{sec:future-work}

We envision \COMETS to be an evolving collection, with additional judgments and tasks added in the future, i.e. knowledge grounded generation, passage ranking, and entity linking.
The topics could also be further enhanced with facet annotations and semantic annotations to support tail and non-KG entities research.

\section{Acknowledgements}
\label{sec:ack}

The authors would like to thank all the domain experts who gave up their valuable time to help develop \COMETS, especially Jamie Macfarlane and Louise Lu. 
Additionally, we'd like to acknowledge Federico Rossetto for his support and help with visualisation.  
This work is supported by the 2019 Bloomberg Data Science Research Grant, the Engineering and Physical Sciences Research Council grant EP/V025708/1, and the 2019 Google Research Grant.

\bibliographystyle{ACM-Reference-Format}
\balance
%\vfill\eject 
\bibliography{foo}

\end{document}